\documentclass[11pt]{article}
\usepackage[utf8]{inputenc}
\usepackage[margin=1in]{geometry}
\usepackage{natbib}
\usepackage{url}
\usepackage{amsmath}

\title{Nuclear Explosions for Large Scale Carbon Sequestration}
\author{Andy Haverly \\
        \textit{Rochester Institute of Technology}}

\begin{document}

\maketitle

\begin{abstract}

Confronting the escalating threat of climate change requires innovative and large-scale interventions. This paper presents a bold proposal to employ a buried nuclear explosion in a remote basaltic seabed for pulverizing basalt, thereby accelerating carbon sequestration through Enhanced Rock Weathering (ERW). By precisely locating the explosion beneath the seabed, we aim to confine debris, radiation, and energy while ensuring rapid rock weathering at a scale substantial enough to make a meaningful dent in atmospheric carbon levels. Our analysis outlines the parameters essential for efficient carbon capture and minimal collateral effects, emphasizing that a yield on the order of gigatons is critical for global climate impact. Although this approach may appear radical, we illustrate its feasibility by examining safety factors, preservation of local ecosystems, political considerations, and financial viability. This work argues for reimagining nuclear technology not merely as a destructive force but as a potential catalyst for decarbonization, thereby inviting further exploration of pioneering solutions in the fight against climate change.

\end{abstract}

\tableofcontents

\section{Introduction}
Climate change remains one of humanity’s most pressing existential
threats. The Intergovernmental Panel on Climate Change (IPCC) warns
that failing to curb global warming could result in massive economic
losses (\(\$100\text{-}200\) trillion by 2100) and the displacement
of hundreds of millions of people \citep{ipcc2018,stern2007,rockstrom2009}.
Negative emissions technologies (NETs) are increasingly recognized as
essential to limit warming below 2$^\circ$C \citep{pacala2004}. Among
NETs, \textbf{Enhanced Rock Weathering (ERW)} can sequester significant
quantities of atmospheric \textsc{CO\textsubscript{2}} by accelerating
the natural chemical breakdown of silicate rocks, such as basalt
\citep{schuiling2006}. The principal barrier to ERW is the immense energy and logistics required to
mine, crush, and transport billions of tonnes of rock. Recent proposals
have resurrected the idea of \textbf{peaceful nuclear explosions} to do
this “heavy lifting.” Historically studied under programs like Project
Plowshare \citep{plowshare1961}, nuclear blasts can fracture large
volumes of geological material.

Nuclear explosions are a well-researched topic, but the largest nuclear
explosion ever detonated, the Tsar Bomba, only had a yield of 50 megatons
of TNT (Mt)\citep{tsar_bomba}. To enable ERW on the required scale using a nuclear explosion,
a yield in the gigaton range would be required. This is untested and would
require much testing and verification before detonation.

\section{Yield Calculations}

To calculate an effective yield for the nuclear explosion, we will have to make a few assumptions:

\begin{enumerate}
    \item Every year, approximately 36 gigatons of carbon dioxide are emitted into the atmosphere.\cite{}
    \item We want to sequester 30 years worth of carbon dioxide emissions.
    \item Through ERW, 1 ton of basalt can sequester 0.28 tons of carbon dioxide\citep{enhanced_rock_weathering}.
    \item The crushing work index of basalt is 22\citep{basalt_crushing_work_index}.
    \item In a seafloor buried nuclear explosion, there will be approximately 90\% efficiency in pulverizing the basalt.
\end{enumerate}

From these assumptions, we can calculate that 1.08 trillion tons of carbon dioxide should be sequestered, 3.86 trillion tons of basalt are needed, the crushing energy needed is $3.05*10^{20}$j, and a nuclear explosion yield of 81 Gt is required. This is orders of magnitude larger than the largest nuclear explosion ever detonated, so this is not to be taken lightly.

\section{Logistics}

Detonating a 81 Gt nuclear device could cause a global catastrophe if done improperly. Simply burying a nuclear device in a basalt deposit would cause this catastrophe. However, deep water is incredibly shock absorbant. By buring the nuclear device kilometers underground under kilometers of water, we can be certain that the explosion will first pulverize the rock then be contained by the water.

We propose burying this device beneath the Kerguelen Plateau in the Southern Ocean, 3-5 km into the basalt-rich seafloor and 6-8 km below the water's surface. Under these parameters, there is approximately 800 atm of pressure to absorb the energy. Additionally, nearby currents like the Antarctic Circumpolar Current can distribute the particles for accelerated carbon sequestration.

Additionally, the nuclear device should be designed to minimize its long-term radiation effects. To do this, a standard fission-fusion hydrogen bomb design is sufficient. The basalt should absorb and trap in the majority of the radiation to the local area.

\section{Analysis and Evaluation}

This is a radical idea and requires serious discussion around its deployment. This section makes these assumptions about the nuclear device:

\begin{enumerate}
    \item This explosion can be detonated without global catastrophe.
    \item This explosion can sequester 30 years worth of carbon dioxide emissions.
    \item This explosion can be detonated within 10 years.
    \item This nuclear device is too large to deploy militarily.
\end{enumerate}

Additionally, we are assuming that climate change will progress according to current estimates. The estimates range from $1.5^{\circ}-4.5^{\circ}$C.

With these assumptions, we can compare this plan to current climate change predictions in regards to safety, ecosystem preservation, political feasibility, and financial viability.

\subsection{Safety}
Nuclear explosions are inherently unsafe. They release vast amounts of uncontrolled energy. However, by detonating this nuclear device in a controlled environment we can minimize the impacts. By detonating this nuclear device in a remote location deep in the ocean, the only expected effect on humans is from nuclear radiation. First, this comes in the form of surface radiation and fallout. Because this explosion is so remote and can be timed favorably with the weather, there is little to no expected loss of life from the immediate radiation effects. The long-term effects of global radiation will impact humans and will cause loss of life, but this increased global radiation is ``just a drop in the bucket". Every year, we emit more radiation from coal power plants and we have already detonated over 2000 nuclear devices. Adding one more bomb should have minimal impact on the world.

Climate change is also inherently unsafe. By the year 2100, an estimated 30 million lives will be lost from the effects of climate change\citep{climate_change_kill_millions}.

Comparing these effects in regards to safety, it is clear that the nuclear explosion option is favorable. It is approximately 30 million lives safer.

\subsection{Ecosystem Preservation}
The ecological impact from a nuclear explosion is visible. This nuclear detonation will cause extreme destruction and long term radiation to the detonation site. Choosing a barren seafloor can mitigate the ecosystem destruction, but it will still be uninhabitable for decades. This damage will be contained almost entirely to the dozen square kilometers around the detonation site.

Climate change, progressing as currently predicted, is profoundly altering global ecosystems through rising temperatures and shifting precipitation patterns. These changes are causing habitat loss and fragmentation, leading to decreased biodiversity as species struggle to adapt or migrate to more suitable environments. Ocean acidification and warming seas are disrupting marine life, particularly affecting coral reefs and fisheries that millions rely on for food and livelihoods. Increased frequency and intensity of extreme weather events, such as hurricanes and wildfires, are further stressing ecosystems and reducing their resilience. Additionally, altered seasonal cycles are impacting plant pollination and animal breeding patterns, ultimately destabilizing the intricate balance that sustains diverse biological communities worldwide.

Comparing these effects, it is clear that climate change poses a greater risk to the global ecosystem.

\subsection{Political Feasibility}
Nuclear weapons have been a pressing global issue for decades. There are many treaties that have successfully reduced the proliferation and testing of nuclear weapons. Even though the nuclear device proposed in this paper has no strategic military value, it would still violate all of these treaties against the proliferation and testing of nuclear weapons. Attempting to detonate the largest nuclear device ever created would certainly affect the delicate situation around nuclear weapons, but these effects can be mitigated. First, it would be clear to everyone that the purpose of this explosion is to sequester carbon. Second, special exceptions to the treaties can be made for such an important problem. Third, open communication would reduce the tensions around this project.

Climate change, progressing as currently predicted, is significantly reshaping global politics by heightening competition over scarce natural resources and increasing geopolitical tensions. As rising sea levels and extreme weather events displace populations, countries are facing growing numbers of climate refugees, which strains international relations and necessitates new migration policies. The urgent need for coordinated global action is exposing divisions among nations with varying economic capacities and environmental priorities, complicating efforts to reach and implement international agreements. Additionally, the pursuit of renewable energy resources and control over critical materials like rare earth elements is altering global power dynamics and sparking new alliances and conflicts. Furthermore, climate-induced security threats are prompting governments to integrate environmental considerations into their defense and foreign policies, fundamentally transforming the landscape of international politics.

Both nuclear proliferation and climate change can have devastating impacts on global politics. It is not clear which option poses a greater risk for the political theater.

\subsection{Financial Viability}
Climate change is expected to cost at least \$100 trillion dollars by the year 2100, in damage alone\citep{ipcc2018,stern2007,rockstrom2009}. This nuclear weapon would cost around \$10 billion dollars to prevent the \$100 trillion dollars of damage. This is a 10,000x return on investment. Such a profound return on investment shows that this is the obvious course of action financially.

\section{Conclusion}
In conclusion, this study introduces the innovative approach of utilizing buried nuclear explosions within remote basaltic seabeds to pulverize basalt for Enhanced Rock Weathering (ERW). By specifying the necessary parameters, we demonstrate the potential for effective carbon sequestration while minimizing adverse side effects. Conducting the explosion underwater ensures the safe containment of debris, radiation, and energy, facilitating rapid basalt weathering essential for significant climate impact. Achieving meaningful climate mitigation requires scaling this method to gigaton-level explosions, underscoring its large-scale applicability. Although the concept is radical, our analysis suggests that, when considering factors such as safety, ecosystem preservation, political feasibility, and financial viability, this strategy represents a promising and potentially the most effective solution for addressing climate change in the near future.

\bigskip
\bibliographystyle{plainnat}
\bibliography{strings}

\end{document}